\documentclass[12pt]{article}
\usepackage[dvips]{graphicx}
\newcommand{\beq}{\begin{equation}}
\newcommand{\eeq}{\end{equation}}
\newcommand{\Frac}[2]{\frac{\displaystyle #1}{\displaystyle #2}}
\newcommand{\lsim}{\stackrel{<}{_\sim}}

\begin{document}
\thispagestyle{empty}
\begin{titlepage}
\begin{center}
\vspace*{-1cm}
\hfill IFIC/01$-$02 \\
\hfill FTUV/01$-$0117
\vspace*{2cm} \\
{\Large \bf The vector form factor of the pion from unitarity}
\vspace*{0.4cm} \\
{\Large \bf and analyticity: a model--independent approach}
\vspace*{1.9cm} \\
{ \sc  A.\ Pich} and {\sc J.\ Portol\'es}
\vspace*{0.8cm} \\
Departament de F\'{\i}sica Te\`orica, IFIC,
CSIC -- Universitat de Val\`encia, \\
Edifici d'Instituts d'Investigaci\'o, Apt.\ Correus
22085, E--46071 Val\`encia, Spain \\ \hfill
\vspace*{1.7cm}
\begin{abstract}
We study a model--independent parameterization of the vector pion
form factor that arises from the constraints of analyticity and 
unitarity. Our description should be suitable up to 
$\sqrt{s} \simeq 1.2 \, \mbox{GeV}$ and allows a model--independent
determination of the mass of the $\rho(770)$ resonance, 
$M_{\rho} = (775.1 \pm 0.5) \, \mbox{MeV}$. We analyse
the experimental data on $\tau^- \rightarrow \pi^- \pi^0 \nu_{\tau}$,
in this framework, and its consequences on the low--energy observables
worked out by chiral perturbation theory. An evaluation of the two pion
contribution to the anomalous magnetic moment of the muon, $a_{\mu}$, and 
to the fine structure constant, $\alpha (M_Z^2)$, is
also performed.
\end{abstract}

\vfill
\end{center}
\end{titlepage}
\newpage
\pagenumbering{arabic}

\section{Introduction}
\hspace*{0.5cm} The hadronic matrix elements of Quantum Chromodynamics
(QCD) currents play a basic role in the understanding of electroweak
processes at the low--energy regime (typically $E \sim 1 \, \mbox{GeV}$). 
However our poor knowledge of the QCD dynamics at these energies 
introduces annoying and serious incertitudes in the description and
prediction of the processes involved.
\par
To bypass this problem several
procedures have been addressed in the literature on this topic. On one
side there is a widespread set of models that pretend to describe, 
in a simplified way, the involved dynamics \cite{GOS,KS}.
 While of importance to 
get a feeling of the entangled physics, the included
simplifying assumptions are usually poorly justified and, sometimes,
even inconsistent with QCD. Ad hoc parameterizations of the matrix 
elements have also been extensively used \cite{KS,TAU}.
The problem with this technique
is that, while the description of data can be properly accounted for, it
is not easy to work out the physics hidden in the parameters. 
\par
A more promising and model--independent procedure is the use of effective
actions from QCD. At very low energies ($E \ll M_{\rho}$, with $M_{\rho}$ the
mass of the $\rho(770)$ resonance) the most important QCD feature is
its chiral symmetry that is realized in
chiral perturbation theory ($\chi$PT) \cite{GL85}, a perturbative
quantum field theory that provides the effective action of QCD in terms
of the lightest pseudoscalar mesons. $\chi$PT has a long and successful
set of predictions both in strong and electroweak processes \cite{PI95}.
At higher energies ($E \sim M_{\rho}$), resonance chiral theory is the
analogous framework \cite{RCH89} where the lightest resonance fields are 
kept as explicit degrees of freedom. With the addition of dynamical
constraints coming from short--distance QCD, resonance chiral theory
becomes a predictive model--independent approach to work with.
\par
One of the simplest hadronic matrix elements of a QCD current is the vector
pion form factor $F_V(s)$ defined through,
\begin{equation}
\langle \, \pi^+ (p) \, \pi^-(p')| \, V_{\mu}^3 \, | \, 0 \, \rangle \; = 
\; (p-p')_{\mu} \, F_V(s) \; \;,
\label{eq:pff}
\end{equation}
where $s = q^2 = (p+p')^2$ and $V_{\mu}^3$ is the third component of the 
vector current associated with the approximate $SU(3)_V$ flavour symmetry of
the QCD
lagrangian. The vector pion form factor drives the hadronic part of both
$e^+ e^- \rightarrow \pi^+ \pi^-$ and $\tau^- \rightarrow \pi^- \pi^0
\nu_{\tau}$ processes in the isospin limit \footnote{If isospin symmetry
is broken, there is a mixing between the third and eighth components of 
the vector current. The spectral
functions are then slightly different in $e^+e^-$ annihilation and tau decays.}. 
There
is an extensive bibliography on the study of this form factor that we do
not review in detail here.
\par
At very low energies, $F_V(s)$ has been calculated in $\chi$PT up to
${\cal O}(p^6)$ \cite{GLPF1,GLPF2}. A successful study at the $\rho(770)$
energy
scale has been carried out in the framework of the resonance chiral
theory (the effective action of QCD at the resonance region) 
in Ref.~\cite{PT97}. In this last reference the unitarity and 
analyticity properties of the vector pion form factor were implemented
in order to match the low--energy result
at ${\cal O}(p^4)$ in $\chi$PT with the correct behaviour at the
$\rho(770)$ peak. The result is in excellent agreement with the data
coming from $e^+ e^- \rightarrow \pi^+ \pi^-$ and $e^- \pi^{\pm} \rightarrow
e^- \pi^{\pm}$ processes. This solution, that includes the $\rho(770)$ 
contribution only, leaves just one free parameter, $M_{\rho}$, and 
provides a suitable description of $F_V(s)$ up to $\sqrt{s} \sim 1 \,
\mbox{GeV}$. If we want to be able to extend its validity at higher
energies we should take into account other contributions. To achieve
this feature, the analyticity and unitarity properties of $F_V(s)$, together
with the resonance chiral theory, continue to provide a model--independent
solution for the vector pion form factor that we analyse, in detail,
in this article. The new solution 
includes two, a priori, unknown parameters in addition to $M_{\rho}$. These
parameters happen to be related to the chiral low--energy observables in 
Ref. \cite{GLPF1,GLPF2}, the squared charged pion radius,
$\langle r^2 \rangle_V^{\pi}$,
and the ${\cal O}(s^2)$ term in the chiral expansion $c_V^{\pi}$.
\par
In the next section we construct the vector pion form factor on grounds
of its analyticity and unitarity relations. In Section 3 we study the
experimental data on $\tau^- \rightarrow \pi^- \pi^0 \nu_{\tau}$ with our
solution for the pion form factor. By a fitting procedure we determine
the values of $M_{\rho}$ and low--energy parameters that tau decay
data demand. Section 4 is devoted to analyse the results we have got from
the fitting procedure and the consequences on the chiral observables of
$\chi$PT. A corresponding evaluation of the two--pion contribution to the
hadronic part of the anomalous magnetic moment of the muon and the 
fine structure constant is collected in Section 5.
We present our conclusions in Section 6.

\section{Analyticity and unitarity in $F_V(s)$}
\hspace*{0.5cm} The vector pion form factor $F_V(s)$
is an analytic function in the whole complex s--plane,
but for the cut along the positive real axis, starting at the lowest
threshold $s = 4 m_{\pi}^2$, where its imaginary part develops a 
discontinuity. This is given by the unitarity condition
\begin{equation}
\mbox{Im} \, F_V(s) \, = \, \Frac{1}{2} \, \sum_n \, \int \, d \rho_n \, 
\langle \, \pi^+ \pi^- \, | \, T^{\dagger} \, | \, n \, \rangle \, 
\langle \, n \, | \, V_{\mu}^3 \, | \, 0 \, \rangle \; \; ,
\label{eq:unit}
\end{equation}
where $|n\rangle$ represents on--shell intermediate states
and $T^{\dagger}$ is the scattering operator connecting
the intermediate state $|n\rangle$ to the final two--pion state.
The first allowed intermediate states are $2 \pi$, $ 4 \pi$ and
$K \overline{K}$. To every intermediate state corresponds a
branch point at the value of $s$ equal to the squared sum of masses of
the corresponding particles, i.e. $s=(2 m_{\pi})^2$, $s = (4 m_{\pi})^2$,
and so on.  In the elastic region, $s < 16 m_{\pi}^2$, the only intermediate
state considered in Eq.~(\ref{eq:unit}) is the one with $2 \pi$, and
Watson final--state
theorem \cite{WA52} relates the imaginary part of $F_V(s)$ to the partial
wave amplitude $t_1^1(s)$ for $\pi \pi$ elastic scattering with angular 
momentum and isospin equal to one. Thus, from Eq.~(\ref{eq:unit}),
\begin{equation}
\mbox{Im} \, F_V(s+i \varepsilon) \, = \,  \sigma_{\pi} \, t_1^1(s) \,
 F_V(s)^* \, = \,
e^{i \, \delta_1^1} \, \sin \delta_1^1 \, F_V(s)^* \; \; ,
\label{eq:unit2}
\end{equation}
where $\sigma_{\pi} = \sqrt{1-4 m_{\pi}^2/s}$. As $\mbox{Im}
F_V(s+i \varepsilon)$
is a real quantity, the phase of $F_V(s)$ must be $\delta_1^1(s)$, that is, 
the phase--shift of the $t_1^1(s)$ partial wave amplitude. Therefore,
\begin{equation}
\mbox{Im} \, F_V(s+i \varepsilon) \, = \, \tan \delta_1^1 \, \mbox{Re} \,
F_V(s) \; \;.
\label{eq:unit3}
\end{equation}
The analyticity and unitarity properties of $F_V(s)$ are accomplished by
demanding that the form factor should satisfy a n--subtracted dispersion
relation in the form
\begin{equation}
F_V(s) \; = \; \sum_{k=0}^{n-1} \, \Frac{s^k}{k!} \, \Frac{d^k}{d s^k} \, 
F_V(s) |_{s=0} \, + \, \Frac{s^n}{\pi} \, \int_{4 m_{\pi}^2}^{\infty} \,
\Frac{dz}{z^n} \, \Frac{\tan \delta_1^1(z) \, \mbox{Re} F_V(z)}{z - s -
i \varepsilon} \; \; ,
\label{eq:drpf}
\end{equation}
where we have used Eq.~(\ref{eq:unit3}). This integral equation has the
known Omn\`es solution \cite{PT97,OM58}
\begin{equation}
F_V(s) \, = \, Q_n(s) \, \exp \left\{ \, \Frac{s^n}{\pi} \, 
\int_{4 m_{\pi}^2}^{\infty} \, \Frac{dz}{z^n} \, 
\Frac{\delta_1^1(z)}{z - s - i \varepsilon} \, \right\} \; \; ,
\label{eq:omsol}
\end{equation}
with
\begin{equation}
Q_n(s) \, = \, \exp \left\{ \, \sum_{k=0}^{n-1} \, \Frac{s^k}{k!} \, 
\Frac{d^k}{ds^k} \, \ln F_V(s) |_{s=0} \, \right\} \; \; .
\label{eq:qns}
\end{equation}
Strictly speaking the solution (\ref{eq:omsol}) for $F_V(s)$ is valid only
below the inelastic threshold $(s < 16 m_{\pi}^2)$. This is because we have
only included the two--pion threshold in the unitarity relation 
(\ref{eq:unit}). However, the contributions from higher multiplicity 
intermediate states are suppressed by phase space and ordinary chiral
counting.
\par
As in any subtracted dispersion relation like the one given by 
Eq.~(\ref{eq:drpf}) there is an interplay between the subtraction constants
(polynomial part) and the dispersive integral. By increasing the number of
subtractions (correspondingly increasing the power of $z$ in the denominator)
we pull in the low--energy part of $\mbox{Im} F_V(s)$ in the
integrand . Then the values
of $\mbox{Im} F_V(s)$ in the upper part of the integration are less 
important. At the same time the information of this high energy region shifts
to the increasing number of subtraction constants that are related 
with the low--energy expansion of the form factor. This situation is 
reflected in the solution of the integral equation (\ref{eq:omsol}). If we
know the $\delta_1^1(s)$ phase--shift only at very low energies, an 
accurate evaluation of the integral in Eq.~(\ref{eq:omsol}) would require
a high number of subtractions. This exchange of information between high
and low energies is, by no means, paradoxical. It is a strict consequence
of the fact that, being an analytic function in the complex s--plane, the
behaviour of $F_V(s)$ at different energy scales is related. Dispersion
relations embody rigorously this property.
\par
The $\delta_1^1(s)$ phase--shift is rather well known, experimentally,
up to $E \sim 2 \, \mbox{GeV}$. Resonance chiral theory provides a
model--independent analytic expression that describes properly the
$\rho(770)$ contribution \cite{PT97} to it~:
\begin{equation}
\delta_1^1(s) \, = \, \arctan \left\{ \, 
\Frac{M_{\rho} \, \Gamma_{\rho}(s)}{M_{\rho}^2 - s} \, \right\} \; \; ,
\label{eq:d11rc}
\end{equation}
with $\Gamma_{\rho}(s)$ the hadronic off--shell $\rho$ width \cite{DTY} (see
Eq.~(\ref{eq:apb2}) in the Appendix). This
result, that provides our definition of $M_{\rho}$, follows from 
Eq.~(\ref{eq:unit3}) and the expression for $F_V(s)$ obtained in 
Ref.~\cite{PT97} that we collect in Appendix A.
The description of data given by $\delta_1^1(s)$ in Eq.~(\ref{eq:d11rc})
is accurate enough
up to $E \sim 1 \, \mbox{GeV}$ for values of $M_{\rho}$ in the ballpark
of the average value collected in the Review of Particle Properties \cite{PDG}.
At higher energies heavier resonances with the
same quantum numbers pop up and to get a correct description we should 
use the available experimental data from Ochs \cite{OC73}.
\par
We will take the result for $F_V(s)$ in Eq.~(\ref{eq:omsol}) with 3 
subtractions. There are several reasons to take this case. On one side
the number of subtractions is high enough to weight the low--energy 
behaviour of $\delta_1^1(s)$ that is much well known than its high energy
part. On the other side the number of subtraction constants, 
three a priori unknown
parameters, is low enough to allow a reasonable parameterization. In fact
one of the subtraction constants is provided by the normalization condition
on the form factor, i.e. $F_V(0) = 1$, and there remain two parameters that can 
be related to the low--energy expansion of the form factor,
 $\langle r^2 \rangle^{\pi}_V$ and $c^{\pi}_V$, as we will shortly see.
\par
Therefore we take as the vector pion form factor provided by analyticity
and unitarity the expression
\begin{equation}
F_V(s) \, = \, \exp \left\{ \, \alpha_1 \, s \, + \, \Frac{1}{2} \,
\alpha_2 \, s^2 \, + \, \Frac{s^3}{\pi} \, \int_{4 m_{\pi}^2}^{\Lambda^2} \,
\Frac{dz}{z^3} \, \Frac{\delta_1^1(z)}{z - s - i \varepsilon} \, 
\right\} \; \; .
\label{eq:3sub}
\end{equation}
Since Eq.~(\ref{eq:unit3}) is only valid in the elastic region,
we have introduced an upper cut in the integration, $\Lambda$. This cut--off
has to be taken high enough not to spoil the, a priori, infinite interval
of integration, but low enough that the integrand is well known in the
interval. As commented above we know best $\delta_1^1(s)$ 
up to $E < 2 \, \mbox{GeV}$. We will take $\Lambda = 2.0 \, \mbox{GeV}$ though,
with three subtractions, there is a negligible difference (within the errors)
between $\Lambda = 1.5 \, \mbox{GeV}$ and the previous value. 
\par
The two subtraction constants $\alpha_1$ and $\alpha_2$ are related with
the squared charge radius of the pion $\langle r^2 \rangle^{\pi}_V$ and the
quadratic term $c^{\pi}_V$ in the low--energy expansion of the pion form
factor
\begin{equation}
F_V(s) \, = \, 1 \, + \, \Frac{1}{6} \, \langle r^2 \rangle^{\pi}_V \, s \, 
+ \, c^{\pi}_V \, s^2 \, + \, {\cal O}(s^3) \; \; ,
\label{eq:chir}
\end{equation}
through the relations
\begin{eqnarray}
\langle r^2 \rangle^{\pi}_V \, & = & \, 6 \, \alpha_1 \; \; , \nonumber \\
c^{\pi}_V \, & = & \, \Frac{1}{2} \, ( \, \alpha_2 \, + \, \alpha_1^2 \, )
\; \; ,
\label{eq:compar}
\end{eqnarray}
that follow from the expansion of the form factor in Eq.~(\ref{eq:3sub}) and
its comparison with Eq.~(\ref{eq:chir}). We will use them to predict
these observables.

\section{The mass of the $\rho(770)$ resonance from a fit to 
$\tau$ decay data}
\hspace*{0.5cm} The fact that $F_V(s)$  is dominated by the $\rho(770)$
vector meson up to $E \sim 1 \, \mbox{GeV}$ has been extensively used to 
get the properties of this resonance. In order to proceed, a 
Breit--Wigner--like form factor is usually introduced and fitted to the
data. This procedure, however, relies in a modelization of the form 
factor that is not necessarily consistent with QCD. Here we propose a
thorough model--independent determination of the mass of the $\rho(770)$
resonance, $M_{\rho}$, defined by Eq.~(\ref{eq:d11rc}).
\par
$F_V(s)$ endows the hadronic dynamics in the $\tau^- \rightarrow \pi^- 
\pi^0 \nu_{\tau}$ decay and the $e^+ e^- \rightarrow \pi^+ \pi^-$ 
process. The experimental data from this last source \cite{BA85,AM86} 
has been available for long time and deeply analysed. The decay  
$\tau^- \rightarrow \pi^- \pi^0 \nu_{\tau}$ has recently been
measured accurately, in the energy region of our interest, by three 
experimental groups~: ALEPH \cite{EXP}, CLEO--II \cite{EXP1} and OPAL
\cite{EXP2}. 
We take $F_V(s)$, as given by Eq.~(\ref{eq:3sub}), to fit the 
ALEPH set of data. 
\par
An appropriate study of the form factor requires a proper description of 
the $\delta_1^1(s)$ phase--shift in the integration interval. As we are
working with 3 subtractions the main contribution to the integration
in Eq.~(\ref{eq:3sub}) comes from the low--energy region of the phase--shift.
However if we wish to consider $F_V(s)$ around $\sqrt{s} \sim 1 \, \mbox{GeV}$
the cut--off $\Lambda$ should be not lower than, let us say, 
$\sqrt{s} \simeq 1.5 \, \mbox{GeV}$, as we commented previously. Therefore
we require a precise description of $\delta_1^1(s)$ in this energy region.
We achieve this through the following procedure~: $\delta_1^1(s)$ given
by Eq.~(\ref{eq:d11rc}) provides an implementation up to 
$\sqrt{s_{match}} = M_{\rho}$;
hence for $M_{\rho} \leq \sqrt{s} \lsim 1.5 \, \mbox{GeV}$ (higher values of
$\sqrt{s}$ being unimportant because the three subtractions performed) 
we include the Ochs set of data \cite{OC73}. As a result we come out with 
a description of $\delta_1^1(s)$, in the region of interest, that contains
all the necessary physics input.
\par
However there are still contributions to the form factor in 
Eq.~(\ref{eq:3sub}) that are not taken into account with Ochs data. These
are those of coupled channels that open at the $K \overline{K}$ threshold
\cite{OOP}. Therefore, in order to have a conservative determination of the
observables, we choose to fit ALEPH data in the range 
$0.32 \, \mbox{GeV} \lsim \sqrt{s} \lsim 1.1 \, \mbox{GeV}$ where
we have a thorough
control of the contributions. The fitting procedure is carried out with the
MINUIT package \cite{MIN}. We find
\begin{eqnarray}
M_{\rho} \, & = & ( \, 775.13 \, \pm \, 0.02 \, ) \, \mbox{MeV} \; \; \; , 
\nonumber \\
\alpha_1 \, & = & \, ( \, 1.84 \, \pm \, 0.02 \, ) \, \mbox{GeV}^{-2} 
\; \; \; , \nonumber \\
\alpha_2 \, & = & \, ( \, 4.18 \, \pm \, 0.05 \, ) \, \mbox{GeV}^{-4}
\; \; \; , \nonumber \\
\chi^2 / d.o.f. \, & = & \,  33.8 / 21 \; \; \; .
\label{eq:res2}
\end{eqnarray}
Though the $\chi^2/d.o.f.$ value found can be considered reasonable it
is necessary to notice that $80 \%$ of $\chi^2$ comes from just three
points \footnote{One of them at $\sqrt{s} \simeq 0.70 \, \mbox{GeV}$ and
the other two around $\sqrt{s} \simeq 0.85 \, \mbox{GeV}$.}.
Errors in Eq.~(\ref{eq:res2}), given by the MINUIT program, are to be taken
with care. They do not include those that come from the choices we have
made in our approach: the energy range to be fitted, number of subtractions,
upper cut of integration $\Lambda$ and the matching point, 
$\sqrt{s_{match}}$, between Ochs 
data and Eq.~(\ref{eq:d11rc}). We estimate the final errors
by exploring the stability of the results with two and four subtractions,
varying the cut--off from $\Lambda = 1.5 \, \mbox{GeV}$ to 
$\Lambda = 2.0 \, \mbox{GeV}$, extending the fitted energy range
up to $\sqrt{s} \simeq 1.6 \, \mbox{GeV}$ and shifting 
$\sqrt{s_{match}}$ within the Ochs data errors. Hence we conclude the figures
\begin{eqnarray}
\; & \; & \nonumber \\ 
M_{\rho} \, & = & (\,775.1 \, \pm \, 0.5 \,) \, \mbox{MeV} \; \; \; , 
\nonumber \\
\alpha_1 \, & = & \, ( \, 1.84 \, \pm \, 0.05 \, ) \, \mbox{GeV}^{-2} 
\; \; \; , \nonumber \\
\alpha_2 \, & = & \, ( \, 4.2 \, \pm \, 0.2 \,  ) \, \mbox{GeV}^{-4}
\; \; \; .
\label{eq:res2fi}
\end{eqnarray}
The parameters $\alpha_1$ and $\alpha_2$ turn out to be 
highly anti--correlated. This procedure provides a mass for the
$\rho(770)$ resonance roughly 5 standard deviations higher
than the Particle Data Group new average \cite{PDG} that is 
$M_{\rho} = (769.3 \pm 0.8) \, \mbox{MeV}$ but consistent with
their average from $\tau$ decays and $e^+ e^-$ processes, 
$M_{\rho} = (776.0 \pm 0.9) \, \mbox{MeV}$. 

\begin{figure}[!tbh]
\begin{center}
\includegraphics[angle=-90,width=0.9\textwidth]{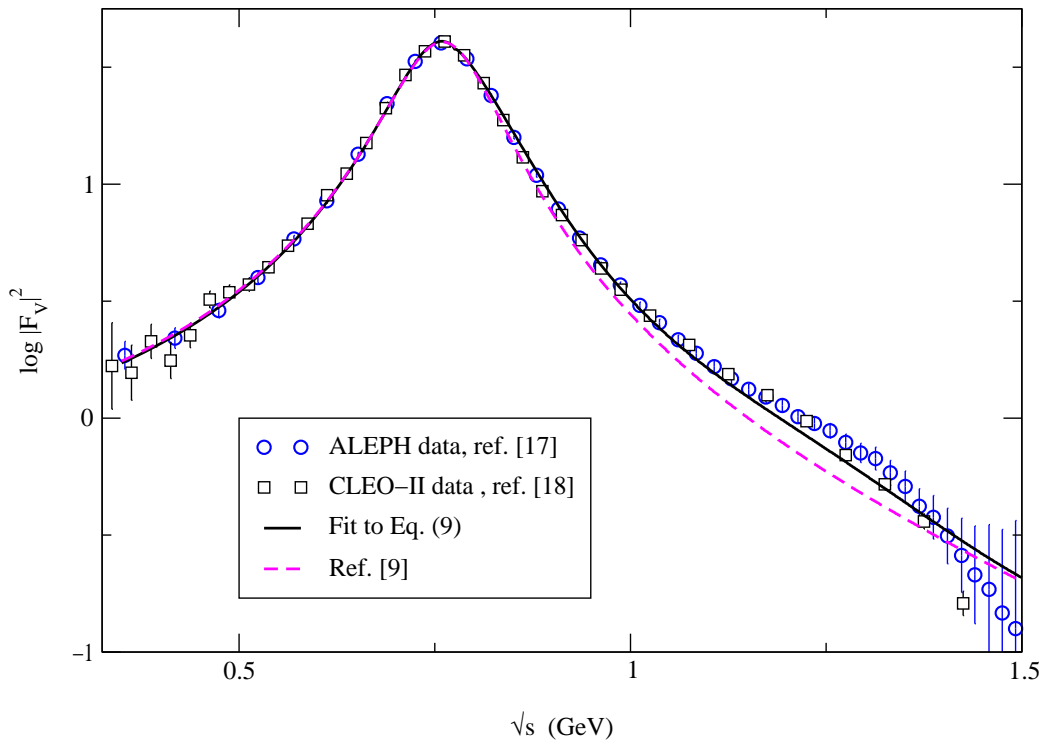}
\end{center}
\caption[]{\label{fig1} \it Comparison of the result of the fit to ALEPH
data with the experimental ALEPH \protect{\cite{EXP}} and CLEO-II 
\protect{\cite{EXP1}} data on $F_V(s)$ from 
$\tau^- \rightarrow \pi^- \pi^0 \nu_{\tau}$
in the $\rho(770)$ energy region. The result of Ref. \protect{\cite{PT97}}
for $M_{\rho} = 775 \, \mbox{MeV}$ is also shown. Up to 
$\sqrt{s} \sim 0.8 \, \mbox{GeV}$ both curves are almost indistinguishable.}
\end{figure}
\begin{figure}[!tb]
\begin{center}
\includegraphics[angle=-90,width=0.9\textwidth]{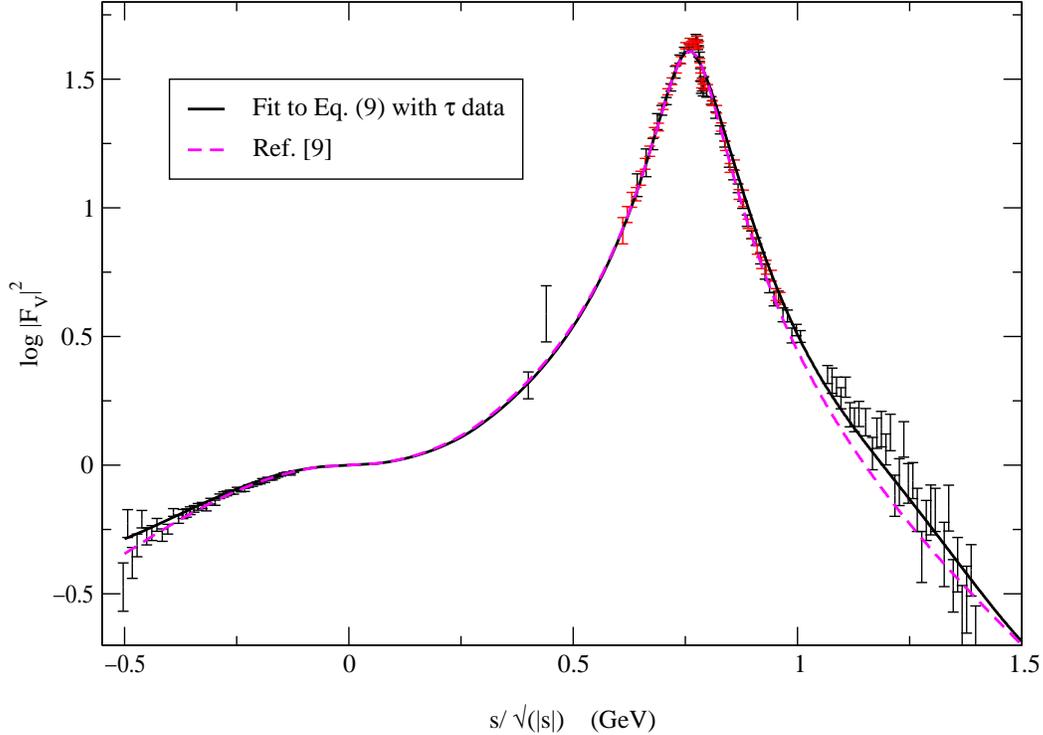}
\end{center}
\caption[]{\label{fig2} \it Comparison of the result of our fit with the
experimental data on $F_V(s)$ from $e^+ e^- \rightarrow \pi^+ \pi^-$
 (time--like) 
\protect{\cite{BA85}} and $e^- \pi^{\pm} \rightarrow e^- \pi^{\pm}$
(space--like) \protect{\cite{AM86}}. The result of Ref. \protect{\cite{PT97}}
($M_{\rho} = 775 \, \mbox{MeV}$) is also shown.
 In the region $-0.4 \, \mbox{GeV} \lsim \, s/\sqrt{|s|} \,
 \lsim 0.8 \, \mbox{GeV}$ both curves are almost indistinguishable.}
\end{figure}

In Fig. 1 we compare the experimental data with our prescription. We also
include the parameter--free prediction (1 subtraction only) of 
Ref. \cite{PT97} that, for completeness, we recall in Appendix A. It can
be seen that our fit gives a good description of data up to 
$E \sim 1.3 \, \mbox{GeV}$. 
Experimental data (in spite of the big errors 
in the higher energy region) seem to have a determinate structure
(mild shoulder) around $E \sim 1.3 \, \mbox{GeV}$.
This could be due to a heavier $\rho$--like resonance as the $\rho(1450)$. 
Our solution takes into account this possibility though, because Ochs data
embody these resonances up to $E \sim 1.5 \, \mbox{GeV}$.
\par
We can compare the results of our fit to tau decay data with the experimental
results 
coming from $e^+ e^- \rightarrow \pi^+ \pi^-$ (time--like) \cite{BA85}
and $e^- \pi^{\pm} \rightarrow e^- \pi^{\pm}$ (space--like) \cite{AM86}
processes. In Fig. 2 we show these sets of data together with the same
curves of Fig. 1. We conclude that the agreement of our fit with data is
good within the errors. Notice that $e^+ e^- \rightarrow \pi^+ \pi^-$ 
data has a contribution from $\omega(782)$ that translates into a 
slight deformation on the right--hand side of the $\rho(770)$ peak. This
is due to a small $I=0$ component contributing to the $2 \pi$ spectral
function in $e^+ e^- \rightarrow \pi^+ \pi^-$. This contribution does
not appear in the isovector spectral function from $\tau^- \rightarrow
\pi^- \pi^0 \nu_{\tau}$ which we are describing.

\section{The low--energy observables}
\hspace*{0.5cm} At $E < 2 m_{\pi}$ the vector pion form factor satisfies
a low--energy expansion given by Eq.~(\ref{eq:chir}). Up to the quadratic
term in $s$ we have, therefore, two low--energy observables, the squared
charge radius of the pion, $\langle r^2 \rangle^{\pi}_V$, and the
quadratic term $c^{\pi}_V$, that are related with the parameters $\alpha_1$
and $\alpha_2$ of the form factor (\ref{eq:3sub}) as given in 
Eq.~(\ref{eq:compar}).
\par
$\langle r^2 \rangle^{\pi}_V$ and $c^{\pi}_V$ have recently been determined
at ${\cal O}(p^6)$ in $\chi$PT \cite{GLPF2}. While chiral symmetry
constraints successfully provide
the chiral logarithms, it remains an incertitude in the polynomial part
that involves counterterms not predicted by the chiral framework. Therefore
it is not possible to give a plain prediction for these observables. 
The authors of Ref.~\cite{GLPF2} performed, by including properly the
chiral logarithms, a fit of the pion form factor, as given by ${\cal O}(p^6)$
$\chi$PT, to the data from $\tau^- \rightarrow \pi^- \pi^0 \nu_{\tau}$,
$e^+ e^- \rightarrow
\pi^+ \pi^-$ and $e^- \pi^{\pm} \rightarrow e^- \pi^{\pm}$ in the low--energy
region ($E \lsim 0.5 \, \mbox{GeV}$). Our procedure provides the low--energy
observables from a fit to a larger energy interval in the time--like
region. In Table 1
we compare our figures with those of Ref.~\cite{GLPF2}.
As can be seen the results compare very well but the errors to the
observables provided by our procedure are smaller (noticeably in $c_V^{\pi}$).

\begin{table}
\begin{center}
\begin{tabular}{|c|c|c|} 
\hline
\multicolumn{1}{|c|}{} &
\multicolumn{1}{|c|}{$\langle r^2 \rangle^{\pi}_V \, (\mbox{GeV}^{-2})$} &
\multicolumn{1}{|c|}{$c^{\pi}_V \, (\mbox{GeV}^{-4})$} \\
\hline
\hline
& & \\
Our fit &  $11.04 \pm 0.30$ &  $3.79 \pm 0.04$ \\
& & \\
\hline
& & \\
${\cal O}(p^6)$ $\chi$PT  &  $11.22 \pm 0.41$ &  $3.85 \pm 0.60$ \\
& & \\
\hline
\end{tabular} 
\caption{\it Low--energy observables of the vector pion form factor up to the
quadratic term. We give the results for our fit and the
${\cal O}(p^6)$ $\chi$PT analysis of Ref.\protect{\cite{GLPF2}}.}
\end{center}
\end{table}
\begin{table}
\begin{center}
\begin{tabular}{|c|c|c|} 
\hline
\multicolumn{1}{|c|}{} &
\multicolumn{1}{|c|}{$r^V_{V_1}(M_{\rho}) \times 10^3$} &
\multicolumn{1}{|c|}{$r^V_{V_2}(M_{\rho}) \times 10^4$} \\
\hline
\hline
& & \\
Our fit &  $-0.79 \pm 0.19$ &  $1.46 \pm 0.03$ \\
& & \\
\hline
& & \\
${\cal O}(p^6)$ $\chi$PT & $-0.68 \pm 0.26$ & $ 1.50 \pm 0.44$ \\
& & \\
\hline
& & \\
VMD  &  $-0.25$ &  $2.6$ \\
& & \\
\hline
\end{tabular} 
\caption{\it Combination of ${\cal O}(p^6)$ counterterms appearing
in the $\chi$PT evaluation of $\langle r^2 \rangle^{\pi}_V$ and
$c^{\pi}_V$.
We give the predictions from our fit and the
ones from the chiral fit and the VMD result of Ref.\protect{\cite{GLPF2}}.}
\end{center}
\end{table}

As commented above the predictability of $\chi$PT at ${\cal O}(p^6)$ is
spoiled because chiral symmetry does not provide information on the 
finite part of the counterterms in the results of 
$\langle r^2 \rangle^{\pi}_V$ and $c^{\pi}_V$. Two combinations of 
${\cal O}(p^6)$ counterterms, $r^r_{V_1}(M_{\rho})$ and 
$r^r_{V_2}(M_{\rho})$, one on each observable, have to be considered.
In order to predict these terms one has to rely in modelizations or
dynamical assumptions like Vector Meson Dominance (VMD). 
This last resource was employed in 
Ref.~\cite{GLPF2} to evaluate the vector resonance contributions
$r_{V_1}^V(M_{\rho})$, $r_{V_2}^V(M_{\rho})$ that is the dominant
piece by far.  
\par
Numerically the ${\cal O}(p^6)$ $\chi$PT expressions relating the
low--energy observables with the
polynomial terms are \footnote{For a complete discussion see 
Ref.~\cite{GLPF2}. We take for $\langle r^2 \rangle^{\pi}_V$ their 
Set I possibility. Our numbers differ slightly from the ones given
in that reference because we use for the pion decay constant
$F_{\pi} = 92.4 \, \mbox{MeV}$ instead of $F_{\pi} = 93.2 \, \mbox{MeV}$.
We neglect the small local contribution from pseudoscalars.}
\begin{eqnarray}
\langle r^2 \rangle^{\pi}_V \, & = & \, \left[ \, 12.312 \, + \, 
1603.4 \, r^V_{V_1}(M_{\rho}) \, \right] \, (\mbox{GeV}^{-2}) \; \; ,
\nonumber \\
c^{\pi}_V \, & = & \, \left[ \, 1.787 \, + \, 13718.7 \, r^V_{V_2}(M_{\rho})
\, \right] \, (\mbox{GeV}^{-4}) \; \; .
\label{eq:rrr}
\end{eqnarray}
Within VMD these counterterms are obtained by integrating out vector
resonances using the resonance chiral theory framework \cite{RCH89}. They
have been worked out, within the Proca formalism, in Ref.~\cite{GLPF2}
with the results
\begin{eqnarray}
r_{V_1}^V \, & = & \, 2 \sqrt{2} \Frac{F_{\pi}^2}{M_V^2} \, f_{\chi} f_V
\; \; , \nonumber \\
r_{V_2}^V \, & = & \, \Frac{F_{\pi}^2}{M_V^2} \, g_{V} f_V
\; \; , 
\label{eq:rvs}
\end{eqnarray}
obtained by integrating the lightest octet of vector resonances of mass
$M_V$. The couplings $f_V$, $g_V$ and $f_{\chi}$ can be phenomenologically
obtained from $\rho \rightarrow e^+ e^-$, $\rho \rightarrow \pi^+ \pi^-$
and $\phi \rightarrow K \overline{K}$ with the results: $f_V=0.20$, 
$g_V= 0.09$ and $f_{\chi}= -0.03$, and, therefore, giving values for
$r_{V_i}^V$ that we collect in Table 2. We compare these VMD results with 
the ones obtained from our fit and the ones provided by the ${\cal O}(p^6)$ 
$\chi$PT fit. We notice that the result
of VMD seems to undervalue $|r^V_{V_1}(M_{\rho})|$ and overestimates
$r^V_{V_2}(M_{\rho})$. As can be seen from Eq.~(\ref{eq:rrr}) this
difference would affect most the value of $c^{\pi}_V$. It has to be 
observed though that, on one side, to extract $r^V_{V_1}(M_{\rho})$ from
$ \langle r^2 \rangle^{\pi}_V$ in Eq.~(\ref{eq:rrr}) a strong
cancellation driven by the term $(\langle r^2 \rangle^{\pi}_V - 12.312)$
is involved
and therefore it is very sensitive to the value of the squared charge
radius of the pion (this problem does not arise in the $r^V_{V_2}(M_{\rho})$
case); 
on the other side, VMD can only offer a rough estimate because,
at this order, heavier resonances could also give a noticeable
contribution while the VMD result only includes the lightest octet of
vector mesons. By neglecting these heavier states we could invert the 
procedure and use our fit to predict the products of couplings 
$f_{\chi}f_V$ and $g_V f_V$ from Eq.~(\ref{eq:rvs}). We obtain, for 
example, $f_{\chi}/g_V = (-1.9 \pm 0.6)$ far from the
phenomenological value $f_{\chi}/g_V \simeq -0.33$. It looks as if the
role of heavier resonances is crucial in order to describe ${\cal O}(p^6)$
vector driven contributions in $\chi$PT.

\section{Two--pion contribution to the muon $(g-2)$ and to $\alpha(M_Z^2)$}
\hspace*{0.5cm} The hadronic contribution to the anomalous magnetic moment
$a_{\mu} = (g_{\mu}-2)/2$ of the muon is the main source of incertitude in its
theoretical prediction. Its leading part comes from the
photon vacuum polarization
insertion into the electromagnetic vertex of the muon. It gives
\cite{DH98}~:
\begin{equation}
a_{\mu}^{had} (vac. \, pol.) \, = \, (692.4 \pm 6.2) \times 10^{-10} \; \; .
\label{eq:amuh}
\end{equation}
This contribution
can be evaluated in terms of the experimental hadronic total cross--section
$\sigma (e^+ e^- \rightarrow \mbox{hadrons})$, where 
$e^+ e^- \rightarrow \pi^+ \pi^-$ is, by far, the dominant part at low 
energies. The bulk, both of the central value ($\sim 75 \%$) and the error
($\sim 80 \%$), of $a_{\mu}^{{had}}$ in Eq.~(\ref{eq:amuh}) comes from 
this $\pi \pi$ intermediate state \cite{ADH98}.
\par
The relevant dispersion integral to evaluate this contribution is (up to 
two loops) \cite{GR69}
\begin{eqnarray}
a_{\mu}^{\pi \pi} \; & = & \; \left( \Frac{\alpha(0) \,  m_{\mu}}{3 \, \pi}
\right)^2 \, \int_{4 m_{\pi}^2}^{\infty} \, \Frac{ds}{s^2} \, 
R_{\pi \pi}(s) \, \hat{K}(s) \; \; \; , \nonumber \\
& & \label{eq:amin} \\
R_{\pi \pi}(s) \; & = & \; \Frac{3 \, s}{4 \, \pi \, \alpha^2(s)} \, 
\sigma(e^+ e^- \rightarrow \pi^+ \pi^-) \; \; , \nonumber
\end{eqnarray}
where the $\hat{K}(s)$ function is given in Ref.~\cite{EJ95}. In terms
of $F_V(s)$ we have
\begin{equation}
a_{\mu}^{\pi \pi} \; = \; \left( \Frac{\alpha(0) \, m_{\mu}}{6 \, \pi}
\right)^2 \, 
\int_{4 m_{\pi}^2}^{\Lambda^2} \, \Frac{ds}{s^2} \, \sigma_{\pi}^3 \, 
|F_V(s)|^2 \, \hat{K}(s) \; \; ,
\label{eq:amfv}
\end{equation}
where we have introduced a cut--off $\Lambda$ as the upper limit of 
integration. As $\hat{K}(s)$ grows
mildly at high values of $s$, the integration in $a_{\mu}^{\pi \pi}$ in 
Eq.~(\ref{eq:amfv}) is dominated by the very low--energy region that gives
the main contribution.
\par
The hadronic contribution to the shift in the fine
structure constant $\Delta \alpha(s)$, defined through 
$\alpha(s) = \alpha(0)/(1-\Delta \alpha(s))$,
can be evaluated from $e^+ e^- \rightarrow \mbox{\em hadrons}$ data by using
a dispersion relation together with the optical theorem \cite{CG60}.
The last estimation has been worked out in Ref.~\cite{DH98} giving
\begin{equation}
\Delta \alpha^{(5)} (M_Z^2) |_{had} \, = \, (276.3  \pm 1.6 ) \times 10^{-4} \;
\; ,
\label{eq:deltanu}
\end{equation}
where the superscript indicates that only the 5 lightest quark flavours
have been considered.
\par
The $\pi \pi$ contribution can be accounted for by
\begin{equation}
\Delta \alpha(M_Z^2) |_{\pi \pi} \; = \; - \, 
\Frac{\alpha(0) \, M_Z^2}{12 \, \pi} \, \int_{4 m_{\pi}^2}^{\Lambda^2} \, 
ds \, \Frac{\sigma_{\pi}^3 \, |F_V(s)|^2}{s (s-M_Z^2)} \; \; ,
\label{eq:delal}
\end{equation}
where, once more, we have introduced a cut--off $\Lambda$ as the upper
limit of integration in order to control the good description of the
integrand. Contrarily to what happens in the $a_{\mu}^{\pi \pi}$ case,
from Eq.~(\ref{eq:delal}) we see that the integrand is not so 
dominated by the low--energy region and, therefore,  higher energy  
contributions are relevant to evaluate 
$\Delta \alpha(M_Z^2) |_{\pi \pi}$. In addition, and as we will see, the 
$\pi \pi$ contribution to 
$\Delta \alpha(M_Z^2) |_{had}$ in this energy region is just a 
modest $10 \%$ of the full value (\ref{eq:deltanu}).

\begin{table}
\begin{center}
\begin{tabular}{|c|c|c|} 
\hline
\multicolumn{1}{|c|}{$\Lambda$ (GeV)} &
\multicolumn{1}{|c|}{$a_{\mu}^{\pi \pi} \times 10^{10}$} &
\multicolumn{1}{|c|}{$\Delta \alpha(M_Z^2)|_{\pi \pi} \times 10^4$} \\
\hline
\hline
& & \\
1.0 &  $505 \pm 6$ &  $33.8 \pm 0.4$ \\
& & \\
\hline
& & \\
1.1  & $511 \pm 6$ & $ 34.7 \pm 0.5$ \\
& & \\
\hline
& & \\
1.2 &  $514 \pm 6$ &  $35.1 \pm 0.5$ \\
& & \\
\hline
& & \\
1.3 & $516 \pm 6$ & $35.4 \pm 0.5$ \\
& & \\
\hline
\end{tabular} 
\caption{\it Values of $a_{\mu}^{\pi \pi}$ and $\Delta \alpha(M_Z^2)
|_{\pi \pi}$ given by our fit to ALEPH $\tau$ decay data, in the
whole energy range ($0.32 \, \mbox{GeV} \leq \sqrt{s} \leq 1.6 \, 
\mbox{GeV}$), for different values of the $\Lambda$ cut--off.}
\end{center}
\end{table}

The study on the vector form factor of the pion that we have carried out
allows us to put forward a prediction for both $a_{\mu}^{\pi \pi}$ and 
$\Delta \alpha (M_Z^2) |_{\pi \pi}$ that we work out as follows. The fit
to ALEPH data that gave our results in Eq.~(\ref{eq:res2fi}) was limited
to $\sqrt{s} \leq 1.1 \, \mbox{GeV}$. As commented there we took this 
region because we have a thorough control of the physics involved within.
At higher energies new physics input, unaccounted for, appears. As a result,
in Fig. 1 it can be seen that our fit misses barely the data above 
$\sqrt{s} \sim 1.2 \, \mbox{GeV}$, well outside the fitted region. The
computation of the integrals in $a_{\mu}^{\pi \pi}$ (\ref{eq:amfv}) and
$\Delta \alpha (M_Z^2) |_{\pi \pi}$ (\ref{eq:delal}) requires a good
knowledge of $F_V(s)$ up to $s \simeq \Lambda^2$, therefore if we wish
to reach $\Lambda \simeq 1.3 \, \mbox{GeV}$ we would need a better 
description of data than the one given with the parameters in 
Eq.~(\ref{eq:res2fi}). To achieve this feature we fix 
$M_{\rho} = 775.1 \, \mbox{MeV}$, as concluded in Eq.~(\ref{eq:res2fi}), 
and leave $\alpha_1$, $\alpha_2$ as free parameters. Then we fit the
ALEPH data in the whole range 
$0.32 \, \mbox{GeV} \leq \sqrt{s} \leq 1.6 \, \mbox{GeV}$.
By studying, as above, the stability of the fitted parameters against
variations in the number of subtractions, the upper limit $\Lambda$, and
the matching point $\sqrt{s_{match}}$, we conclude the values
$\widetilde{\alpha_1} = (1.83 \pm 0.03) \, \mbox{GeV}^{-2}$,
$\widetilde{\alpha_2} = (4.28 \pm 0.08) \, \mbox{GeV}^{-4}$,
consistent with the solution of the restricted fit (\ref{eq:res2fi}) but
with smaller errors.
The tildes on $\alpha_1$ and $\alpha_2$ are meant to prevent their
use in Eq.~(\ref{eq:compar}). We emphasize that $\widetilde{\alpha_1}$ and
$\widetilde{\alpha_2}$ are not proper physical values of the $\alpha_1$,
$\alpha_2$ parameters because we have
fitted a region of experimental data that is not properly implemented
theoretically. However the above values of $\widetilde{\alpha_1}$ and
$\widetilde{\alpha_2}$ describe well data up to 
$\sqrt{s} \simeq 1.3 \, \mbox{GeV}$ and, therefore, are useful to evaluate
the integrals in $a_{\mu}^{\pi \pi}$ and $\Delta \alpha (M_Z^2) |_{\pi \pi}$
with smaller errors. The values we get are collected in Table 3.
\par
It has to be noticed that our errors are similar to those
obtained in recent estimations \cite{ADH98}, though the results in
this reference were obtained from a combination of 
$e^+ e^- \rightarrow \pi \pi$ and $\tau^- \rightarrow \pi^- \pi^0 \nu_{\tau}$
decay data while our results come from a fit to this last process up to 
$\sqrt{s} \simeq 1.6 \, \mbox{GeV}$. An improvement on our errors would 
require an analysis of the pion vector form factor
 with a more complete
set of data, combining $e^+ e^- \rightarrow \pi \pi$ and
$\tau^- \rightarrow \pi^- \pi^0 \nu_{\tau}$
processes.

\section{Conclusions}
\hspace*{0.5cm} To gain access to the resonance properties, from experimental 
data, a correct definition of those properties has to be theoretically
implemented. The use of modelizations, though sometimes unavoidable,
can spoil seriously the conclusions obtained from data. In this article we
have studied the vector pion form factor $F_V(s)$ within a model--independent 
approach. We have introduced a parameterization of the form factor
provided by the 
all--important properties of its analyticity and unitarity relations. 
This last construction relates $F_V(s)$ to the $\delta_1^1(s)$ phase--shift
of elastic $\pi \pi$ scattering.
\par
To proceed we have included the $\delta_1^1(s)$ phase--shift (up to 
$\sqrt{s} \simeq 1.5 \, \mbox{GeV}$) with a model--independent parameterization,
provided by the resonance chiral theory and experimental data. Our form
factor depends on two, a priori unknown, subtraction constants and the
$\rho(770)$ mass. We have fitted ALEPH data on 
$\tau^- \rightarrow \pi^- \pi^0 \nu_{\tau}$ to the form factor for 
$E \lsim 1.1 \, \mbox{GeV}$ and we obtain $M_{\rho} = (775.1 \pm
0.5) \, \mbox{MeV}$. Our result for $M_{\rho}$ is bigger than the
new average of the Review of Particle Properties \cite{PDG} but very 
much consistent with their average from $\tau$ decays and $e^+ e^-$ annihilation
processes. The predictions given by our results on the low--energy
observables worked out in $\chi$PT, $\langle r^2 \rangle_V^{\pi}$ and
$c_V^{\pi}$ have also been computed. We find good agreement with the results
from the fit in $\chi$PT though our errors are smaller. It is necessary
to notice, though, that when these figures are worked on to determine
local chiral ${\cal O}(p^6)$ counterterms, the values we get are not 
consistent with those obtained, through VMD, from resonance chiral theory
by integrating out the lightest octet of vector resonances. As a conclusion
it seems that room is left for the contribution of heavier
resonances.
\par
Finally we have evaluated the $\pi \pi$ contribution to the anomalous
magnetic moment of the muon, $a_{\mu}^{\pi \pi}$, and the shift of the
fine structure constant $\Delta \alpha(M_Z^2) |_{\pi \pi}$. An improvement
in the theoretical errors of these quantities would be achieved with a more
complete analysis of the available data.
\par
We have shown how it is possible to extract model--independent
information of resonances from experimental data by exploiting general
properties of form factors, such as unitarity and analyticity. When 
combined with the resonance chiral theory, the effective action of QCD
at the lightest resonance region, these properties provide a compelling
framework for the study of form factors.
\vspace*{0.7cm} \\
\noindent {\large \bf Acknowledgements} \\ \\
\hspace*{0.5cm}
We wish to thank Jon Urheim for correspondence on the $\tau$ decay data
from CLEO-II.
This work has been supported in part by TMR, EC--Contract No.\
ERB FMRX-CT98-0169 and by CICYT (Spain) under grant PB97-1261.

\appendix
\newcounter{erasmo}
\renewcommand{\thesection}{\Alph{erasmo}}
\renewcommand{\theequation}{\Alph{erasmo}.\arabic{equation}}
\renewcommand{\thetable}{\Alph{erasmo}}
\setcounter{erasmo}{1}
\setcounter{equation}{0}
\setcounter{table}{0}

\section*{Appendix A}
\hspace*{0.5cm}  A theoretical 
construction of the vector form factor of the pion was performed 
in Ref.~\cite{PT97} by matching
the ${\cal O}(p^4)$ $\chi$PT result (valid at $E \ll M_{\rho}$) with
the prescription provided by the resonance chiral theory. The
procedure also took into account the analyticity and unitarity 
properties of $F_V(s)$. The result only includes the contribution of
the $\rho(770)$ resonance and gives an excellent description
of data up to $E \sim 1 \, \mbox{GeV}$ with just one parameter, 
$M_{\rho}$. We have compared this prescription with ours in Figures
1 and 2.
\par
For completeness we recall here the result of Ref.~\cite{PT97}:
\begin{eqnarray}
F_V(s) \,  =  \, \Frac{M_{\rho}^2}{M_{\rho}^2 \, - \, s \, - \, 
i M_{\rho} \Gamma_{\rho}(s)} \, &  & \! \! \! \! \! \! \! \! 
\exp \left\{ \, \Frac{-s}{96 \pi^2 
F_{\pi}^2} \, \left[ \, \mbox{Re} A \left( m_{\pi}^2/s, m_{\pi}^2/M_{\rho}^2
\right) \, + \right. \right. \nonumber \\
& &   \left. \left. 
\; \; \; \; \; \; \; \; \; \; \; \; \; \; \; \; \; \; 
\, \Frac{1}{2} \, \mbox{Re} 
A \left( m_{K}^2/s, m_{K}^2/M_{\rho}^2 \right) \, \right] \right\} \;  ,
\label{eq:apb1}
\end{eqnarray}
where $\Gamma_{\rho}(s)$ is the hadronic off--shell width of the $\rho(770)$
resonance
\cite{DTY},
\begin{equation}
\Gamma_{\rho}(s) \, = \, \Frac{M_{\rho} s}{96 \pi F_{\pi}^2} \, 
\left[ \sigma_{\pi}^3 \theta(s-4 m_{\pi}^2) \, + \, \frac{1}{2} 
\sigma_K^3 \theta(s-4 m_K^2) \, \right] \; \; ,
\label{eq:apb2}
\end{equation}
and
\begin{equation}
A \left( m_P^2/s, m_P^2/\mu^2 \right) \, =  \, \ln \left( \Frac{m_P^2}{\mu^2}
\right) \, + \, 8 \, \Frac{m_P^2}{s} \, - \, \Frac{5}{3} \, + \, 
\sigma_P^3 \, \ln \left( \Frac{\sigma_P + 1}{\sigma_P - 1} \right) \; \;,
\label{eq:apb3}
\end{equation}
with $\sigma_P = \sqrt{1 - 4 m_P^2/s}$.

\end{document}